\documentclass[aps,prl,twocolumn,superscriptaddress,showpacs]{revtex4}
\usepackage{graphicx}

\begin{document}

\title{Probing spin correlations with phonons in the strongly frustrated
magnet ZnCr$_2$O$_4$}

\author{A. B. Sushkov}
\affiliation{Materials Research Science and Engineering Center,
University of Maryland, College Park, Maryland 20742}

\author{O.~Tchernyshyov}
\affiliation{Department of Physics and Astronomy, The Johns Hopkins
University, Baltimore, Maryland 21218}

\author{W.~Ratcliff II}
\affiliation{Department of Physics and Astronomy, Rutgers University,
Piscataway, New Jersey 08854}

\author{S. W. Cheong}
\affiliation{Department of Physics and Astronomy, Rutgers University,
Piscataway, New Jersey 08854}

\author{H. D. Drew}
\affiliation{Materials Research Science and Engineering Center,
University of Maryland, College Park, Maryland 20742}


\date{\today}

\begin{abstract}
The spin-lattice coupling plays an important role in strongly
frustrated magnets.  In ZnCr$_2$O$_4$, an excellent realization of the
Heisenberg antiferromagnet on the ``pyrochlore'' network, a lattice
distortion relieves the geometrical frustration through a
spin-Peierls-like phase transition at $T_c = 12.5$~K.  Conversely,
spin correlations strongly influence the elastic properties of a
frustrated magnet.  By using infrared spectroscopy and published data
on magnetic specific heat, we demonstrate that the frequency of an
optical phonon triplet in ZnCr$_2$O$_4$ tracks the nearest-neighbor
spin correlations above $T_c$.  The splitting of the phonon triplet
below $T_c$ provides a way to measure of the spin-Peierls order
parameter.
\end{abstract}

\pacs{
75.50.Ee, 
78.30.Hv,  
75.30.Et, 
75.10.Hk 
}

\maketitle


Geometrically frustrated magnets can resist magnetic ordering and
remain in a strongly correlated paramagnetic state well below the
Curie-Weiss temperature $\Theta_{\rm CW}$
\cite{Schiffer,Ramirez,Greedan}.  Particularly strong frustration is
found in the Heisenberg antiferromagnet on the ``pyrochlore lattice'',
wherein magnetic sites form a lattice of corner-sharing tetrahedra.
The dynamics of the magnet is determined by the spin Hamiltonian
\begin{equation}
H = J \sum_{\langle ij \rangle} {\bf S}_i \cdot {\bf S}_j,
\end{equation}
with the interaction restricted to nearest-neighbor bonds
$\langle ij \rangle$.
Theoretical investigations indicate that classical spins on this
lattice may not order down to zero temperature \cite{Zinkin,Moessner}.
Cubic spinel ZnCr$_2$O$_4$ offers a good realization of this model.
Observation of magnetic order at low temperatures \cite{Oles,Lee-2000}
has been explained in terms of a spin-driven Jahn-Teller effect
\cite{yamashita,Oleg-PRL-2002} that is reminiscent of the spin-Peierls
instability in spin chains.  The coupling between spin and lattice
degrees of freedom thus plays a major role in relieving the
geometrical frustration.

\begin{figure}
\includegraphics[width=0.9\columnwidth]{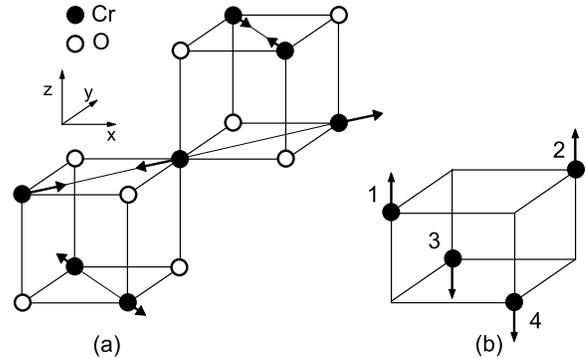}
\caption{(a) One (out of five) of the symmetry coordinates of the
$F_{1u}$ mode which modulates Cr--Cr exchange, arrows show
displacements; (b) spin ordering expected for a tetragonally distorted
tetrahedron in ZnCr$_2$O$_4$, arrows indicate spins. }
\label{cub}
\end{figure}

The B sites of ZnCr$_2$O$_4$ are occupied by magnetic ions Cr$^{3+}$
with spin $S=3/2$ [Fig.~\ref{cub}(a)].  A crystal field of a nearly
cubic symmetry splits the five $3d$ orbitals of Cr$^{3+}$ into a
$t_{2g}$ triplet and an $e_{g}$ doublet.  By Hund's rules, the three
electrons of Cr$^{3+}$ have aligned spins ($S=3/2$) and occupy all of
the three $t_{2g}$ states.  The lack of orbital degeneracy precludes
the ordinary Jahn-Teller effect common to spinels and ensures a very
small spin anisotropy \cite{Lee-2000}.  The shape of the $t_{2g}$
orbitals, pointing towards the neighboring Cr$^{3+}$ ions, makes
direct exchange the primary mechanism for magnetic interactions
\cite{Goodenough,Motida}.  As a result, interactions beyond nearest
neighbors are negligibly small, while the nearest-neighbor
interactions are highly sensitive to the motion of Cr$^{3+}$ ions
creating a strong spin-phonon coupling.

The magnetic susceptibility of ZnCr$_2$O$_4$ follows the Curie-Weiss
law at high temperatures with $\Theta_{CW}=390$~K, which gives the
nearest-neighbor exchange coupling $J = 4.5$ meV \cite{Lee-2000}.  As
the temperature is lowered below $\Theta_{\rm CW}$, the magnet
gradually enters a paramagnetic state with strong correlations between
spins but no magnetic order: the spins remain liquid but their motions
are highly coordinated \cite{Lee-2002}.  At $T_c = 12.5$~K the
magnet undergoes a first-order transition into a phase with antiferromagnetic
order and a structural distortion \cite{Lee-2000}.  According to the
theory \cite{yamashita,Oleg-PRL-2002,Oleg-PRB-2002}, the phase
transition is driven by local distortions of the tetrahedra that have
the $E$ symmetry in the language of the tetrahedral point group $T_d$.
This is consistent with the observation of a tetragonal distortion
below $T_c$ \cite{Lee-2000}.

In this work we demonstrate that lattice vibrations can provide
quantitative information about spin correlations.  We have measured the
frequencies of IR-active phonons and found, below the magnetoelastic
phase transition, a large splitting of a phonon mode involving magnetic
ions.  From the magnitude of the splitting we have inferred the
absolute value of the spin-Peierls order parameter $\langle {\bf S}_i
\cdot {\bf S}_j - {\bf S}_k \cdot {\bf S}_l \rangle$, where $\langle ij
\rangle$ and $\langle kl \rangle$ are nearest-neighbor bonds
\cite{Oleg-PRB-2002}.

{\em Experimental details.}
Powders of ZnCr$_2$O$_4$ were prepared in air using the standard solid
state reaction method. Small single crystals of ZnCr$_2$O$_4$ were
grown using these powders using chemical transport method in quartz
tubes sealed with Cl$_2$ gas as a transport agent. The crystals were of
regular habit with the $\langle 111 \rangle$ face clearly
visible. The working surface of our sample is tilted slightly
($\sim$3.5$^{\circ}$) from the $\langle 111 \rangle$ plane as
determined by X-rays. Reflectivity measurements were performed using a
Fourier-transform spectrometer in the frequency range from 100 to
5000~cm$^{-1}$.  The temperature dependence from 6 to 300~K was
measured using liquid helium in a continuous flow cryostat (sample in
vacuum) with optical access windows.

\begin{figure}
\includegraphics[width=0.9\columnwidth]{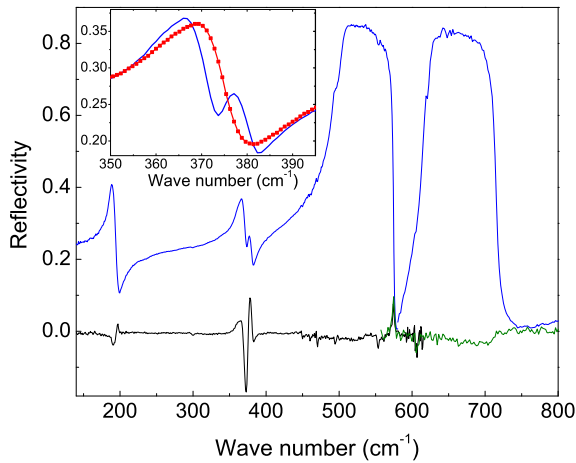}
\caption{(Color online) Top curve: $R_{LT}$ --- reflectivity spectrum
of ZnCr$_2$O$_4$ single crystal in the low-temperature phase; bottom
curve: $2(R_{LT}-R_{HT})$; inset: $R_{LT}$ and $R_{HT}$ (line+symbols)
in a narrow range. } \label{R}
\end{figure}

The room temperature phonon spectrum of cubic spinels is well studied.
Our measurements at room temperature agree with previously published
data for single crystals of ZnCr$_2$O$_4$ \cite{Lutz-X}.  As expected
on symmetry grounds, the spectrum consists of four phonon triplets of
the $F_{1u}$ symmetry (point group $O_h$, space group $Fd\bar{3}m$).
Figure~\ref{R} shows the reflectivity spectrum $R_{\rm LT}$ for the
low-temperature (LT) phase averaged over the temperature range from 9
to 11 K.  It also shows a magnified difference of the spectra $2(R_{\rm
LT}-R_{\rm HT})$, where $R_{\rm HT}$ is the reflectivity spectrum
averaged between 13 and 14 K. The most prominent difference is observed
at 370~cm$^{-1}$: below $T_c$ this phonon splits into two components.
In addition, there is a slight modification of the vibrational mode at
186~cm$^{-1}$, which cannot be fit with two slightly split oscillators.
Several new modes with small oscillator strengths are seen above
300~cm$^{-1}$ in the low-temperature IR spectra.  Most likely, these
are phonons with nonzero wave vectors in the cubic phase that become
visible in the LT phase due to Brillouin zone folding
\cite{Damascelli,Popova2002}.  This is consistent with a recent
observation of an enlarged structural unit cell below $T_c$
\cite{H.Ueda}. Raman spectra also contain a set of new phonons in the
LT phase~\cite{Girsh}.

To extract the temperature dependencies of all phonon parameters, we
fit the reflectivity $R = |\frac{1-\sqrt\epsilon}{1+\sqrt\epsilon}|^2$
using a model dielectric function
\begin{equation}
\epsilon(\omega, T) = \epsilon_\infty+\sum_j
\frac{S_j\omega_{j}^2}{\omega_{j}^2-\omega^2-i\omega\gamma_j}
\label{eq-epsilon}
\end{equation}
where $\epsilon_\infty(T) \equiv\epsilon(\infty,T)$ is the dielectric
constant well above all phonons, $j$ enumerates the phonons,
$S_j(T)$ is an oscillator strength, $\omega_{j}(T)$ is a phonon
frequency, and $\gamma_j(T)$ is a damping rate.

\begin{figure}
\includegraphics[width=0.9\columnwidth]{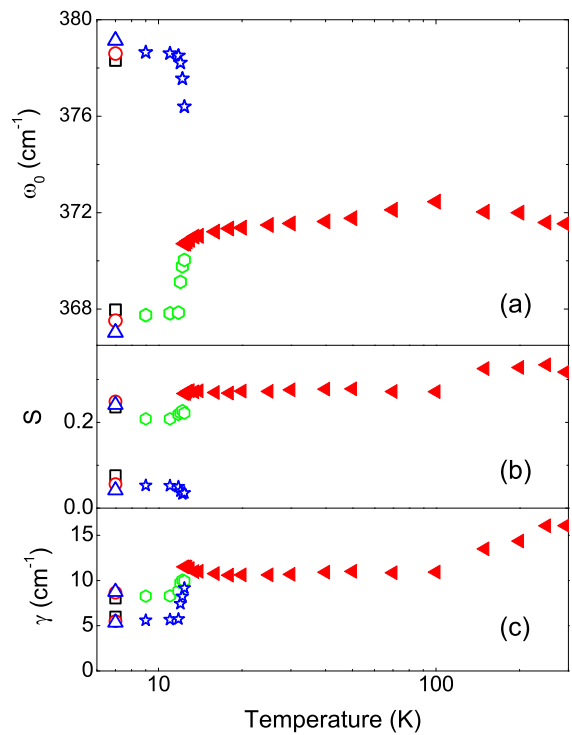}
\caption{(Color online) Temperature dependence of the fit
parameters for the split phonon in ZnCr$_2$O$_4$; symbols --- fit;
squares, circles, and triangles at 7~K --- fits of polarized spectra
with $P=0^{\circ}, 45^{\circ}, 90^{\circ}$, respectively. }
\label{w2}
\end{figure}

Figure~\ref{w2} presents temperature dependences of the phonon
frequency, oscillator strength, and damping rate for the 370~cm$^{-1}$
phonons. The symbols show the best-fit parameters for the model
dielectric function (\ref{eq-epsilon}) with four phonon modes above
$T_c$ and five phonons below.  Upon cooling from room temperature, the
resonance frequency of the 370~cm$^{-1}$ phonon hardens first, as do
the other three IR-active modes.  In contrast, it softens
significantly below 100 K [Fig.~\ref{w2}(a)]. Just below $T_c=12.5$~K,
this phonon splits into two modes with a frequency difference of
11~cm$^{-1}$.  The total oscillator strength is approximately
conserved.  There is no polarization dependence of the reflectivity
spectrum at 14~K.  Some polarization dependence appears on the split
phonon below $T_c$ as can be seen from data points in Fig.~\ref{w2} at
7~K.  A clear observation of the expected polarization effects in the
tetragonal phase was not found most likely because of multiple domains in the
sample. No hysteresis effect is observed.

{\em Theoretical model.}
The splitting of the 370~cm$^{-1}$ phonon is consistent with the
lowering of the crystalline symmetry from cubic to tetragonal.  What
physical mechanism is responsible for the magnitude of splitting?
Anharmonic effects unrelated to magnetism could cause the splitting,
however there are two arguments against this interpretation.  First,
the magnitude of the anharmonic effect can be estimated as $\Delta
\omega /\omega = \gamma \Delta a/a$, where $\gamma$ is an appropriate
Gr\"{u}neisen parameter.  To reconcile the observed splitting $\Delta
\omega /\omega = 0.030$ with the magnitude of the tetragonal
distortion $\Delta a/a \approx 10^{-3}$ \cite{Lee-2000}, we must
assume a rather large Gr\"{u}neisen constant $\gamma \approx 30$,
which must be explained.  Second, the three other IR-active $F_{1u}$
modes observed in the experiment exhibit much smaller
($\Delta\omega\leq 0.2$~cm$^{-1}$) frequency related changes in the
tetragonal phase.  These objections cast doubts on anharmonic (and
nonmagnetic) origins of the splitting.

Most likely, the splitting of the triplet is caused by the same effect
that triggers the spin-Peierls instability --- the spin-phonon coupling.
It is well known that the elastic constants in magnetic materials are
affected by the spins \cite{Baltensperger,Lockwood}.  As a
consequence, the phonon frequency is sensitive to correlations of
spins of nearest-neighbor pairs $\langle ij \rangle$:
\begin{equation}
\omega = \omega_0 + \lambda \langle {\bf S}_i \cdot {\bf S}_j \rangle,
\label{eq-lambda}
\end{equation}
where the constant $\lambda$ has a typical value of a few cm$^{-1}$
when optical phonon frequency is well above all magnons and phonon
is not a zone folding mode;
in addition to taking a thermal average, the spin
correlations must also be averaged over the crystal with the weights
appropriate for a given phonon.

A strong argument in favor of the magnetoelastic mechanism is the
nature of the 370~cm$^{-1}$ phonon: it features by far the largest
contribution of the symmetry coordinate modulating Cr--Cr bonds (Fig.~\ref{cub}a)
among the four IR-active
modes \cite{Himmrich}.  The particular geometry of the occupied $3d$
orbitals in chromium makes the direct Cr--Cr exchange the primary
source of magnetic interactions \cite{Goodenough,Motida}.  In the case
of direct exchange, the strength of magnetic interactions is
particularly sensitive to the distances between the magnetic ions
(rather than the bond angles in the case of superexchange).  Therefore
we expect to find the largest splitting in those modes which produce
the largest modulations of Cr--Cr distances, as indeed observed.

\begin{figure}
\includegraphics[width=0.9\columnwidth]{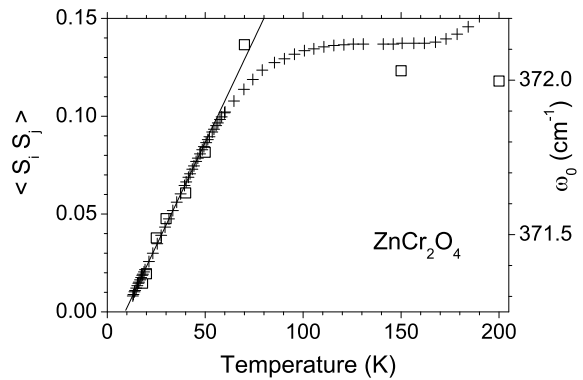}
\caption{Crosses: nearest-neighbor correlations $\langle {\bf S}_i
\cdot {\bf S}_j \rangle$, up to an arbitrary additive constant,
derived from the specific-heat data of Martinho {\em et al.}
\cite{Martinho-PRB-1}; straight line fits crosses below 50~K.
Squares: the phonon frequency.}
\label{fig-calibration}
\end{figure}

To verify the linear relation (\ref{eq-lambda}) and to determine the
proportionality constant $\lambda$, we have compared the temperature
dependences of the spin correlations and phonon frequency in the
high-temperature phase (Fig.~\ref{fig-calibration}).  The
nearest-neighbor spin correlations $\langle {\bf S}_i \cdot {\bf S}_j
\rangle$ were determined from published data of specific heat
\cite{Martinho-PRB-1}.  Neglecting the magnetoelastic effects in the
undistorted phase~\cite{mag-el} we relate the spin correlations to the
magnetic part of the specific heat per mole $C_m$:
\begin{equation}
\langle {\bf S}_i \cdot {\bf S}_j \rangle = \mbox{const}
+ \frac{1}{6N_A J} \int_{T_c}^T C_m(T) \, {\rm d}T,
\label{eq-int}
\end{equation}
where $6 N_A$ is the number of Cr--Cr bonds per mole.  The scaling
relation (\ref{eq-lambda}) works fairly well in the temperature range
between 18 and 150~K. (The phonon softening above 100~K is likely due
to thermal expansion.)  This procedure yields the scaling constant
$\lambda = 6.2 \mbox{ cm}^{-1}$.

The relatively large value of $\lambda$
can be ascribed to the strong sensitivity of
direct exchange to atomic displacements: the exchange constant $J$
varies with the Cr--Cr distance approximately exponentially,
\begin{equation}
J(r+\Delta r) \approx J(r) e^{-\alpha \, \Delta r},
\label{eq-exp}
\end{equation}
on the scale of the Bohr radius $a_B$.  In contrast, the spin-phonon
coupling in the case of superexchange comes from variations of bond
angles, which have a less dramatic effect on magnetic energy.

To obtain a quantitative estimate of $\lambda$, consider a simplified
lattice model in which magnetic ions of mass $m$ are connected by
springs of stiffness $k$ (the rest of the atoms are discarded).  The
resulting ``pyrochlore'' lattice has only one optical mode that
transforms as the irreducible triplet $F_{1u}$ under the operations of
the point group $O_h$.  The spins move along a $\langle 110 \rangle$
direction [Fig.~\ref{cub}(a)] with the frequency $\omega_0 =
2\sqrt{k_0/m}$.  The dependence of the magnetic energy $J(r)({\bf S}_i
\cdot {\bf S}_j)$ on the ion separation $r$ provides a magnetic
contribution to the spring constants $k = k_0 + \delta k_{ex} = k_0 +
J'' \, \langle {\bf S}_i \cdot {\bf S}_j \rangle$, where $J'' = {\rm
d}^2 J(r)/{\rm d}r^2$.  Hence $\lambda = 2J'' / (m\omega_0) \approx
2\alpha^2 J / (m\omega_0)$.

The value of $\alpha$ can be estimated theoretically from the exchange
integral for $3d$ orbitals of hydrogenlike ions, which yields $\alpha
= 2Z/3a_B = 5.0 \mbox{ \AA}^{-1}$ for Cr$^{3+}$ ions ($Z = 4$) and
$\lambda = 3.2 \mbox{ cm}^{-1}$.  An alternative estimate comes from
the experimentally measured variation of the exchange constant with
the distance giving $\alpha = -J'/J = 8.9 \mbox{ \AA}^{-1}$
\cite{Motida,Henning73,Lee-2000} and $\lambda = 10.0 \mbox{ cm}^{-1}$.
Our value $\lambda = 6.2 \mbox{ cm}^{-1}$ is midway between these
estimates.

The observed splitting of the triplet phonon can now be translated
directly into the spin-Peierls order parameter of the distorted phase.
For a tetragonal distortion with lattice constants $a = b \neq c$
(Fig.~\ref{cub}b),
\begin{equation}
\langle {\bf S}_1 \cdot {\bf S}_2
- {\bf S}_2 \cdot {\bf S}_3 \rangle
= (\omega_{z} - \omega_{x})/\lambda = 1.8,
\label{eq-f}
\end{equation}
where bonds 12 are along the direction (110) and bonds 23 are along
(011).  The bond variables $\langle {\bf S}_i \cdot {\bf S}_j \rangle$
are averaged both over the thermal ensemble and over the location in
the crystal.

{\em Discussion.}
The inferred value of the spin-Peierls order parameter (\ref{eq-f}) is
only a fraction of what can be attained in a state with a uniform
tetragonal distortion and collinear spins (Fig.~5 of
Ref.~\onlinecite{Oleg-PRB-2002}), in which bonds along the directions
$(110)$ and $(1\bar{1}0)$ are maximally frustrated: $\langle {\bf S}_1
\cdot {\bf S}_2 \rangle = +S^2$, while the rest of the bonds are
fully satisfied $\langle {\bf S}_1 \cdot {\bf S}_2 \rangle = -S^2$.
In such a state, the $z$ component of the phonon triplet
[Fig.~\ref{cub}(a)] would probe the frustrated bonds with spring
constants $k_0 + J''S^2$, while the remaining two modes would involve
bonds with $k = k_0 - J''S^2$.  We suspect that the primary reason
for this significant reduction of the observed order parameter is the
presence of a substantial nonuniform distortion in the crystal.
Because the spin-Peierls order parameter is linearly coupled to the
crystal distortion \cite{Oleg-PRB-2002}, bond averages $\langle {\bf
S}_i \cdot {\bf S}_j \rangle$ will be nonuniform as well.  Even
though the {\em local} value of the order parameter (\ref{eq-f}) may
be large, averaging over the crystal may reduce it substantially.

Indirect evidence for a nonuniform distortion has been presented by Lee
{\em et al.} \cite{Lee-2000}.  They point out that the observed uniform
tetragonal distortion is too small to account for the magnetic and
elastic energy released in the form of latent heat at the phase
transition.  Direct evidence comes from X-ray measurements of Ueda {\em
et al.} \cite{H.Ueda} and Raman scattering \cite{Girsh}. If indeed the
lion's share of the elastic and magnetic energy is hidden in the
nonuniform component of the lattice distortion, even larger splittings
may be found in phonons with nonzero wave vectors.  For a distortion
with half-integer indices, such as $\langle \frac{1}{2} \frac{1}{2}
\frac{1}{2} \rangle$ \cite{H.Ueda}, the likely candidates are phonon
doublets at the wave vectors $\langle \frac{1}{4} \frac{1}{4}
\frac{1}{4} \rangle$.

In conclusion, we have studied the temperature evolution of optical
phonons of symmetry $F_{1u}$ in a strongly frustrated antiferromagnet
ZnCr$_2$O$_4$.  The changes were most dramatic for the phonon with the
largest amplitudes of Cr--Cr vibrations.  Its frequency showed a marked
softening in the strongly correlated paramagnetic regime below 100~K.
It essentially tracked the temperature evolution of spin correlations
inferred from magnetic specific heat allowing us to calibrate the
frequency for the purposes of measuring spin correlations.  Below $T_c$
the triply degenerate phonon splits into a singlet and a doublet
providing direct information about the uniform part of the spin-Peierls
order parameter (\ref{eq-f}).  Its magnitude is only a fraction of what
could be attained in a state with a uniform tetragonal distortion and
collinear spins described in Ref.~\onlinecite{Oleg-PRB-2002}.  It is
therefore likely that the lattice distortion and the spin-Peierls order
parameter are nonuniform.  The nonuniform component of the distortion
may produce an even larger splitting of a phonon doublet with a wave
vector $\langle \frac{1}{4} \frac{1}{4} \frac{1}{4} \rangle$.

We thank  G.~Blumberg  and C.~L.~Broholm for useful discussions.  This
work was supported in part by the National Science Foundation under
Grants No. DMR-0080008 and DMR-0348679.

\end{document}